\documentclass[aps,prl,twocolumn,superscriptaddress]{revtex4}
\usepackage{graphicx}
\usepackage{mathrsfs}

\begin{document}



\title{The environment of graphene probed by electrostatic force microscopy}



\author{J. Moser}
\email{moser.joel@gmail.com}
\affiliation{CIN2 and CNM Barcelona,
Campus UAB, E-08193 Bellaterra, Spain\\}

\author{A. Verdaguer}
\affiliation{ICN, Campus UAB, E-08193 Bellaterra, Spain\\}

\author{D. Jim\'{e}nez}
\affiliation{Departament d'Enginyeria Electr\`{o}nica, Escola T\`{e}cnica Superior d'Enginyeria, Universitat Aut\`{o}noma de Barcelona,
E-08193 Bellaterra, Spain\\}

\author{A. Barreiro}
\affiliation{CIN2 and CNM Barcelona, Campus UAB, E-08193 Bellaterra,
Spain\\}

\author{A. Bachtold}
\email{adrian.bachtold@cnm.es} \affiliation{CIN2 and CNM Barcelona,
Campus UAB, E-08193 Bellaterra, Spain\\}



\date{}

\begin{abstract}
We employ electrostatic force microscopy to study the electrostatic
environment of graphene sheets prepared with the micro-mechanical
exfoliation technique. We detect the electric dipole of residues
left from the adhesive tape during graphene preparation, as well as
the dipole of water molecules adsorbed on top of graphene. Water
molecules form a dipole layer that can generate an electric field as
large as $\sim 10^{9}~\textrm{V}\cdot \textrm{m}^{-1}$. We expect
that water molecules can significantly modify the electrical
properties of graphene devices.
\end{abstract}


\maketitle

Graphene, a two-dimensional crystal of carbon atoms arranged in a
honeycomb lattice, is among the thinnest objects imaginable
\cite{physworld,natmat}. The structural properties of graphene make
it a system of choice to study the physics of Dirac fermions
\cite{novoselov,kim}; it is also envisioned as a building block for
a novel generation of electronic devices. One inherent technological
difficulty however remains: because all the atoms are at the surface
and are directly exposed to the environment, the electronic
properties are easily affected by unwanted adsorbates.

In this Letter, we employ electrostatic force microscopy (EFM) to
probe the electrostatic environment of graphene sheets. Two
adsorbate species possessing an electric dipole are identified:
water molecules and residues left from the adhesive tape during
fabrication. Water molecules form a dipole layer on top of graphene
that can generate an electric field as large as $\sim
10^{9}~\textrm{V}\cdot \textrm{m}^{-1}$. Tape residues form an
ultra-thin layer nearby the graphene sheets on top of the silicon
oxide substrate, which is not detectable using standard topographic
atomic force microscopy.

\begin{figure}[t]
\includegraphics{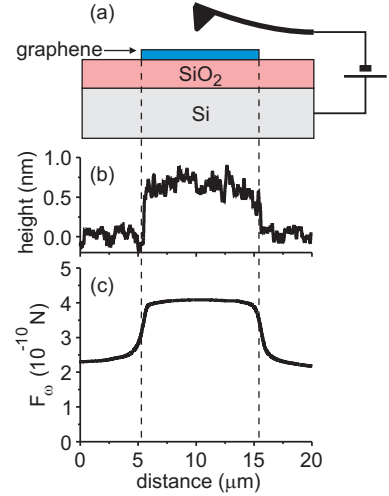}
\caption{(color online) (a) Schematic of the EFM setup. (b) In a
first pass, the topography of the graphene sheet is recorded. (c) In
a second pass, a $dc$ voltage $V_{dc}$ is applied to the Si
substrate and an $ac$ voltage $V_{ac}=1$~mV is applied to the tip at
a frequency set to the resonance frequency of the cantilever. The
sample is scanned at a constant height of about 50~nm and the
oscillating force is measured.}
\end{figure}

We start by briefly describing our fabrication technique and our
experimental setup. Graphene sheets are obtained using the
conventional micro-mechanical exfoliation technique
\cite{electric_field_effect,pnas_geim}: a flake of bulk Kish
graphite is cleaved repeatedly with an adhesive tape and pressed
down onto a silicon wafer coated with 280~nm of thermal silicon
oxide. Two types of adhesive tape are used: standard wafer
protection tape for microfabrication by ICROS and Magic Tape by 3M.
Thin graphene sheets are located optically, imaged by atomic force
microscopy, and occasionally examined by Raman spectroscopy to
identify single layer specimens. EFM measurements are carried out
under various humidity conditions in a constant flow of either dry
${\textrm N_{2}}$ gas, or moist ${\textrm N_{2}}$ gas produced by
bubbling dry ${\textrm N_{2}}$ in deionized water.

The EFM technique \cite{kpfm} is well suited to study dipoles on
surfaces. Our EFM protocol proceeds in two passes along a given scan
line: first the topography is recorded in tapping mode
(Fig.~1(a,b)), then the AFM tip is lifted by a given amount and a
bias $V_{dc}$ is applied to the Si backgate and a potential
$V_{ac}\textrm{cos}(\omega t)$ is applied to the tip. The tip
experiences a force whose term at the frequency $\omega$ reads
(Fig.~1(c)):
\begin{equation}
F_{\omega}=\frac{\partial C}{\partial z}(V_{dc}-\Delta\phi)V_{ac}
\end{equation}
where $C$ and $\Delta\phi$ are the capacitance and the contact
potential difference between the sample and the tip. We measure
$F_{\omega}$ over a range of $V_{dc}$ (Fig.~2(a) and Fig.~3(a-c)).
As in Kelvin probe force microscopy, $\Delta\phi$ is given by
$V_{dc}$ that minimizes $F_{\omega}$.

\begin{figure}[t]
\includegraphics{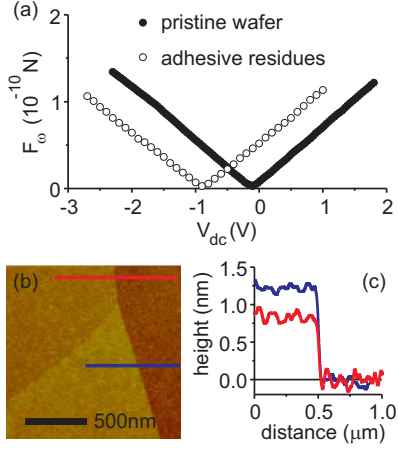}
\caption{(color online) (a) Force term $F_{\omega}$ experienced by
the AFM tip as a function of $V_{dc}$ on a pristine
\textrm{SiO$_{2}$} surface and on \textrm{SiO$_{2}$} covered with
tape residues. (b-c) Topography of a typical graphene sheet. Plots
in (c) are sections along the red and blue lines depicted in (b).}
\end{figure}

Prior to discussing the measurements, we recall the connection
between $\Delta\phi$ and the presence of dipoles on surfaces. The
energy $e\phi$ is the energy barrier that an electron has to
overcome in order to be extracted from a material to the vacuum. Any
electric field existing at the surface of this material, originating
\textsl{e.g.} from dipoles, modifies the contact potential $\phi$ by
an amount $\chi$: $\phi=W+\chi$, where $W$ is the contact potential
without electric field at the surface. For further insight into the
underlying physics of $\Delta\phi$, see
Refs.~\cite{langmuir,ashcroft}. In our experiment, the contact
potential of the tip is $\phi_{tip}$, and the one of the Si wafer
depends on whether the graphene sheet is present or not:
\begin{equation}
\phi_{Si}^{graphene}=W_{Si}+\chi^{graphene}
\end{equation}
\begin{equation}
\phi_{Si}^{no~graphene}=W_{Si}+\chi^{no~graphene}
\end{equation}

We first look into the electrostatic traces left by the adhesive
tape. In the case of a \textrm{SiO$_{2}$} surface onto which the
adhesive tape has been pressed (without graphite), EFM measurements
reveal that the contact potential difference
$\Delta\phi=\phi_{Si}^{no~graphene}-\phi_{tip}$ is significantly
shifted to negative values (Fig.~2(a)). $\Delta\phi$ is found to
vary between -2~V and -0.5~V depending upon where on the wafer the
measurement is carried out. This has to be compared with
$\Delta\phi\simeq 0$~V for a pristine wafer. Both adhesive tapes
that we used yield similar results. This shift in $\Delta\phi$
suggests that adhesive residues change $\chi^{no~graphene}$, which
we attribute to the deposition of dipoles on the surface. Note that
the shift in $\Delta\phi$ is stable over long periods of time, which
suggests that it is not related to individual charges that would get
neutralized, for example, by charged molecules in the environment.

\begin{figure}[t]
\includegraphics{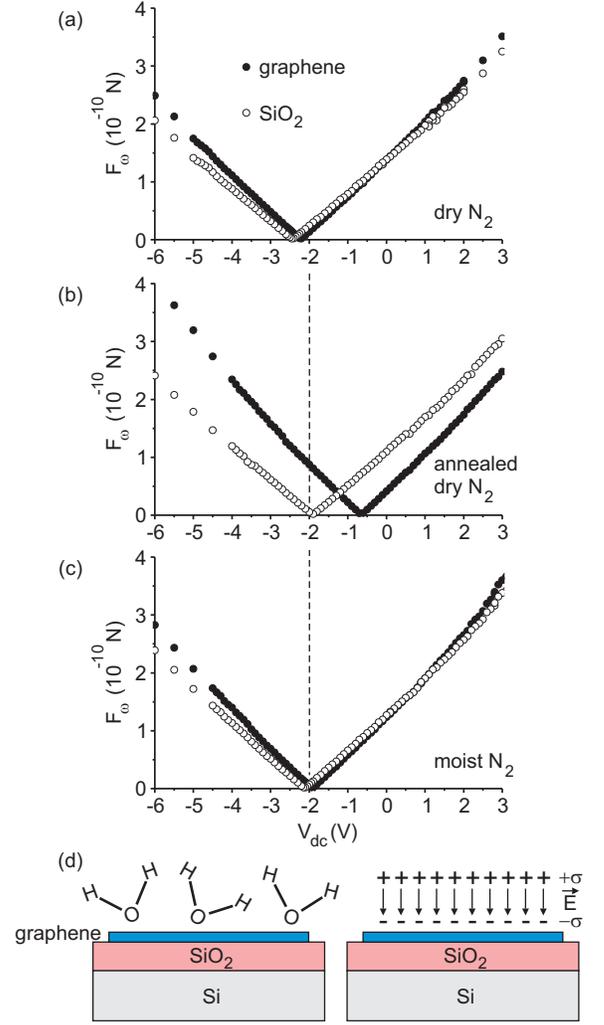}
\caption{(color online) Force term $F_{\omega}$ experienced by the
AFM tip as a function of $V_{dc}$ in (a) dry ${\textrm N_{2}}$; (b)
dry ${\textrm N_{2}}$ after heating the sample to $160^\circ$C for
1~hour in dry ${\textrm N_{2}}$; (c) moist ${\textrm N_{2}}$. Data
on graphene and on silicon oxide $\sim 1\mu$m away from graphene are
shown. The dotted line at $V_{dc}=-2$~V draws the attention to the
rather weak dependence on humidity of $F_{\omega}(V_{dc})$ measured
on the oxide. (d) Water molecules adsorb on average with the oxygen
atom pointing towards the graphene sheet, and form a dipole layer
with an effective surface charge $\sigma$ (schematic not on scale)}
\end{figure}

These tape residues do not seem to roughen the surface, as inferred
from topography measurements (Fig.~2(b,c)). Presumably this is why
they have not been reported so far, as a standard topography scan
fails to detect them. Further studies are needed, especially to find
out whether or not these residues are present underneath the
graphene sheets. At first sight, they should not, as the
\textrm{SiO$_{2}$} surface is masked by the graphene sheets when
pressing the tape down onto the wafer. However, micro-mechanical
exfoliation is rather difficult to control, and residues may well
lie beneath graphene sheets. The latter situation could have
important consequences on the transport properties of graphene
devices (see discussion below for water).

We now turn to EFM measurements on graphene sheets where we modify
the humidity of the environment. Fig.~3 compares measurements on a
single graphene layer and on the oxide $\sim 1\mu$m away. The
measurement is first performed in dry ${\textrm N_{2}}$ with a
relative humidity RH of less than 3\%, thus corresponding to a
submonolayer water coverage on clean \textrm{SiO$_{2}$} (Fig.~3(a))
\cite{albert}. Without taking it out of the dry environment, the
sample is heated to $160^\circ$C for 1~hour and measured again
(Fig.~3(b)). Eventually, moist ${\textrm N_{2}}$ is introduced until
RH$\simeq 50\%$ ($\sim 3$ water monolayers on clean
\textrm{SiO$_{2}$} \cite{albert}), see Fig.~3(c). $\Delta\phi$ on
the graphene sheet is observed to vary between each step
significantly ($-2.1~\textrm{V}\rightarrow
-0.7~\textrm{V}\rightarrow -1.9~\textrm{V}$) \cite{david}. By
contrast, $\Delta\phi$ measured on the oxide stays pretty much
constant ($\sim-2$~V), which suggests that $\phi_{tip}$, $W$, and
$\chi^{no~graphene}$ are not affected much by humidity
\cite{ref_glue}. Therefore the quantity that appears to be sensitive
to water is $\chi^{graphene}$, which changes by $\sim 1.3$~V from
one step to another in Fig.~3. We attribute this strong variation to
water molecules that desorb from and readsorb onto the graphene
sheet \cite{footnote}.

The adsorption of water on graphene may appear surprising, as it is
well established that graphite is highly hydrophobic
\cite{chakarov}. However, it has been reported that water can adsorb
on graphite for RH$\simeq 60\%$ at the surface defects such as steps
\cite{luna}. It has also been reported that water can adsorb on
carbon structures such as carbon nanotubes \cite{dai} and on
self-assembled monolayers (SAM) of carbon chains \cite{SAM}. It has
been argued that the key ingredient in water adsorption on SAM is
the surface roughness. Interestingly, the surface of graphene is
also corrugated as it tends to follow the roughness of the substrate
\cite{stm1,stm2}.

Because $\Delta\phi=\phi_{Si}^{graphene}-\phi_{tip}$ shifts to more
negative values as humidity is raised, water molecules adsorb on
average with the oxygen atoms pointing towards the graphene sheet.
As such, the layer of water molecules can be described as a dipole
layer with the negative charges towards the graphene sheet (see
Fig.~3(d)). The electric field is maximum within the dipole layer,
and vanishes to zero as $\textrm{e}^{-az}$ as the distance $z$ to
the layer is increased (in the particular case where the layer
consists of a periodic array of dipoles that are all pointing in the
same direction, $a$ is the distance between two neighboring
dipoles).

We can obtain a rough estimate for the strength of the electric
field $E$ within the water dipole layer. Assuming that the width of
the dipole layer is $d=1$~nm, we get $E\simeq
\Delta\chi^{graphene}/d\simeq 10^{9}~\textrm{V}\cdot
\textrm{m}^{-1}$, where $\Delta\chi^{graphene}\simeq 1$~V is the
shift in $\chi^{graphene}$ from dry to wet environment. As a
comparison, we calculate $E$ to be of the order of $10^{9}$ to
$10^{10}~\textrm{V}\cdot \textrm{m}^{-1}$ using the water electric
dipole $6.2\cdot 10^{-30}$~Cm, assuming a monolayer of water
molecules that are all pointing in the direction perpendicular to
the surface with a density from 1 to 10~nm$^{-2}$, and modeling the
dipole layer as a parallel plate capacitor. The strength of $E$
changes somewhat when considering that the dipoles are not all
oriented in the same direction or can flip in the field.

An electric field of $10^{9}~\textrm{V}\cdot \textrm{m}^{-1}$ within the dipole layer
corresponds to the strength obtained when applying $\simeq 300$~V between the Si backgate and a graphene device.
It is important to note that the graphene sheet experiences a fraction of
this field only, as it lies outside of the dipole layer. Nevertheless, we
expect that $E$ will significantly shift the Fermi level of
graphene devices \cite{apl}. Moreover, this field
is likely to be inhomogeneous, resulting in puddles of electrons and
holes near the charge neutrality point \cite{yacoby}. Another consequence
is that water can screen charge impurities \cite{dassarma} as well as modify electron-electron interactions.

In conclusion, we show that EFM is a powerful tool for the
characterization of the electrostatic environment of graphene. Water
molecules form a dipole layer on top of graphene that generates a
large electric field. We expect water to have a strong influence on
the transport properties of graphene devices.

Acknowledgements -- We are thankful to E. Hern\'{a}ndez and F.
Guinea for helpful discussions. We are indebted to F. Alsina for his
help with the Raman characterization. This work was supported by an
EURYI grant, FP6-IST-021285-2, and EXPLORA NAN2007-29375-E,
Ministerio de Educación y Ciencia, Spain. D. J. acknowledges
financial support from Ministerio de Educación y Ciencia under
project TEC2006-13731-C02-01/MIC.




\begin{references}
\bibitem{david}
The slope of $F_{\omega}(V_{dc})$ is different on the oxide and on
the graphene sheet. This is attributed to the graphene sheet that
changes the capacitance between the tip and the Si substrate. In
principle, it should be possible to extract the dielectric constant
$\epsilon_{G}$ of the graphene sheet. We numerically solved the
Poisson equation and found that $\partial C/\partial z$ has a very
weak dependence on $\epsilon_{G}$ when $\epsilon_{G}\gtrsim 2$. As a
result, estimating the dielectric constant of graphene is difficult.
\bibitem{ref_glue}
Note that $\chi^{no~graphene}$ varies slightly as humidity is
changed. The size and the sign of the variation are consistent with
previous work on \textrm{SiO$_{2}$} surfaces contaminated with a
submonolayer of carbon \cite{albert}.
\bibitem{footnote}
Note that the wafer was exposed to air during manipulation before
these measurements, at which point water and other molecules were
likely to adsorb on the graphene sheet.
\end{references}
\end{document}